\documentstyle[12pt]{article}
\newcommand{\gsim}
{\mbox{\hspace*{0.1em}\raisebox{.4ex}{$\scriptstyle >$}
\hspace{-0.88em}\raisebox{-.6ex}{$\scriptstyle\sim$}\hspace*{0.05em}}}
\newcommand{\lsim}
{\mbox{\hspace*{0.1em}\raisebox{.4ex}{$\scriptstyle <$}
\hspace{-0.88em}\raisebox{-.6ex}{$\scriptstyle\sim$}\hspace*{0.05em}}}
\newcommand{\sd}{\makebox[0.5ex]{}}

\begin{document}
\begin{center}
QUARKONIUM MASS SPLITTING REVISITED: EFFECTS OF CLOSED MESONIC
CHANNELS\\[0.2cm]
M.Shmatikov\\[0.2cm]
{\it Russian Research Center "Kurchatov Institute",\\
123182 Moscow, Russia}\\[0.2cm]
{\bf Abstract}\\[0.2cm]
\end{center}
Modifications of the mass spectrum the quarkonium induced by its virtual
dissociation into a pair of heavy mesons is considered. Coupling between quark and 
mesonic channels results in noticeable corrections to spin-dependent
mass splitting. In particular, the observable hierarchy of mass splittings
in the $\chi_c$, $\chi_b$ and $\chi{'}_b$ multiplets is reproduced.\\[0.3cm]
Quarkonium is the  
testing ground for various theoretical approaches and, especially, of
potential models. In particular, analysis of the quarkonium mass spectrum is
used for investigating properties  of the confining potential.
Phenomenological potentials were reasonably successful in describing gross
features of the quarkonia spectra: the level spacing and ordering . More
subtle details of the mass spectra, in particular, spin-dependent mass
splitting still remains a challenging problem. Time-honored approach to the
analysis of spin-dependent effects in quarkonia consists in introducing an
interquark {\it QED-type} potential but with more generic vertex structure \cite{st}: 
\begin{equation}
V=\Gamma _i\tilde{V}(q^2)\Gamma _i,  \label{pot1}
\end{equation}
where $\tilde{V}(q^2)$ is the propagator of the exchanged particle and $%
\Gamma _i$ is the coupling vertex (equal e.g. to 1 or $\gamma ^\mu $ in the
case of scalar and vector-type coupling respectively) . In the
nonrelativistic limit the expansion of (\ref{pot1}) in the inverse powers of the
heavy-quark mass yields the sum of central and spin-dependent components 
\begin{equation}
V=V_c+\left[ (\vec{S}_1+\vec{S}_2)\,\vec{L}\right] V_{LO}+S_{12}\cdot
V_T+\left( \vec{S}_1\cdot \vec{S}_2\right) V_{SS}  \label{pot2}
\end{equation}
with $V_{LO},\;V_T,\;V_{SS}$ standing for the spin-orbit, tensor and
spin-spin potentials respectively. In eq.(\ref{pot2}) $\vec{S}_{1,2}$ are
the quarks' spins, $\vec{L}$ is their orbital momentum and $S_{12}=12(\vec{S}%
_1\cdot \hat{r})(\vec{S}_2\cdot \hat{r})-4(\vec{S}_1\cdot \vec{S}_2).$ The
spin-structure of the interquark potential (\ref{pot2}) can be translated in
the mass splitting of the triplet $P$ states as follows: 
\[
\begin{array}{cc}
m(^3P_0)= & \bar{m}-2a-4b \\ 
m(^3P_1)= & \bar{m}-a+2b \\ 
m(^3P_2)= & \bar{m}+a-2b/5
\end{array}
\]
where $\bar{m}$ is the c.o.g. mass of the triplet (weighted with $2J+1$
factor), $a$ and $b$ are the averaged values of the $V_{LO}$ and $V_T$
potential components respectively. The $V_{SS}$ component of the potential
controls  mass splitting between the triplet and singlet $P$ states. The
combined action of spin-orbit and tensor forces is parameterized in the form
of the ratio 
\begin{equation}
R=\frac{m(^3P_2)-m(^3P_1)}{m(^3P_1)-m(^3P_0)}  \label{ratio}
\end{equation}
The masses of the $^3P_J$ charmonium states as determined by Crystal Ball
 \cite{cry1} yield
\begin{equation}
R(\chi _c)=0.478\pm 0.01  \label{rc}
\end{equation}
with corresponding ratios for $1^3P_J$ and $2^3P_J$ states of the $\bar{b}b$%
-quarkonium reading respectively \cite{cry1} 
\begin{equation}
R(\chi _b)=0.664\pm 0.038  \label{rc1}
\end{equation}
and 
\begin{equation}
R(\chi _b^{\prime })=0.576\pm 0.014  \label{rc2}
\end{equation}
Note the hierarchy of $R$'s as given by the measured quarkonia masses: 
\begin{equation}
R^{\rm exp}(\chi _b)\,\gsim\,R^{\rm exp}(\chi _b^{\prime })>R^{\rm exp}(\chi _c)  \label{hie}
\end{equation}
At the same time potential models involving the sum of the  the scalar
(confining) and vector (residual one-gluon exchange) potentials ({\it cf} 
(\ref{pot1}) ) yielded controversial results predicting $R(\chi _b)>R(\chi
_c)$ and $R(\chi _b^{\prime })$ slightly larger or, at best, approximately
equal to $R(\chi _b)$ in the qualitative disagreement with (\ref{hie}) (see 
\cite{set} for the details):
\begin{equation}
R(\chi _b^{\prime })\,\gsim\,R(\chi _b)>R(\chi _c) 
\end{equation}
 Observed disagreement hints at the existence of
additional effects. A solution to the problem was sought  by
introducing some unconventional interactions, in particular, a pseudoscalar
component of the quark-antiquark potential with $\Gamma =\gamma ^5$ in (\ref
{pot1}) \cite{fra}. Another issue might be an account of relativistic
effects which are more pronounced in $\bar{c}c$ as compared to their $\bar{b}%
b$ counterparts. Currently no convincing solution to the problem in the
framework of the potential-based approach seems to exist. The controversy
casts doubts on the universally adopted conclusion about scalar
Lorentz-transformation properties of the confining potential.

In the present paper we consider an alternative mechanism affecting mass-splitting
in the quarkonia systems which might give an insight to the solution of the
challenging problem. It is coupling of quarkonia states to pairs of heavy
quark-mesons $H\bar{H}$. Investigation of the effects emerging from such
coupling was pioneered by \cite{eich}, where a unified description of the
heavy quark-antiquark binding and of the quarkonium coupling to heavy mesons
was given. However, the calculations made based heavily on the specific form
of the confining potential. The latter was assumed to have  Lorentz-vector
transformation properties in disagreement with the current experimental data
on both the spectroscopy of $\bar{c}c$ states and on their decay amplitudes 
\cite{bar}. The latter circumstance substantiates revisiting effects which
emerge from coupling between quarkonia states and mesonic channels.

Coupling of quarkonium ($Q\bar{Q}$) bound states to heavy-meson 
pairs ($H\bar{H}$) results in the mass
shift of the former because of the virtual dissociation 
\begin{equation}
Q\bar{Q}\rightarrow (Q\bar{q})+(\bar{Q}q)\rightarrow Q\bar{Q}  \label{vir}
\end{equation}
The mass shift emerges as a matter of fact due to an interaction between
heavy $Q\bar{Q}$ and (virtual) light $q\bar{q}$ quark pairs. The situation,
however, is somewhat tricky. Indeed, in the {\it adiabatic} approximation
the presence of the light-quark pairs does not affect (at least) the
long-range part of the potential operating between heavy quark and
antiquark. More precisely, its linear form is maintained, and only the
tension is altered. The latter, however, is a phenomenological parameter
determined from the experimental data implying that in the $m_Q\rightarrow
\infty $ limit coupling to mesonic channels does not result in an observable
effect \cite{jaf}. This conclusions refers to the spectrum gross features
controlled by the string tension. At the same time spin-dependent mass splitting 
is scaled by a $1/m_Q$ factor, and non-vanishing effect are possible.

Consider in more detail the mechanism which produces effective mass
splitting of the $P_J$ triplet. The lightest $Q\bar{q}$ states (and their
antiparticles) materialize in the form of pseudoscalar and vector mesons
(denoted hereafter $H$ and $H^{*}$ respectively). Be $H$ and $H^{*}$
mass-degenerate, coupling between $Q\bar{Q}$ and $q\bar{q}$ would not result
in mass splitting of quarkonium states. However, because of the spin $Q-\bar{q}$
interaction the $H^{*}$ vector meson proves to be heavier than its pseudoscalar $H$
counterpart. It implies that the mass shift of the quarkonium
states is scaled by the width of a $H\bar{H}$ continuous spectrum band up to 
the effective mass $\gsim (m_{H^{*}}-m_H)$. The contribution coming from  $H\bar{H}$ pairs 
with masses exceeding $m_H+m_{H^{*}}$ is canceled by similar contribution of 
$\bar{H}H^{*}$ and $\bar{H}^{*}H^{*}$ heavy-meson states.

A quantum-mechanical consideration of the problem allows making some
qualitative conclusions concerning the expected effect. First, quarkonium
states, depending on their spin-parity quantum numbers, couple to different $%
H\bar{H}$ states whereas some of them do not have a $H\bar{H}$ match at all.
Indeed, the $^3P_0$ and $^3P_2$ quarkonium states couple to the $H\bar{H}$
pairs with $J^\pi =0^{+}$ and $2^{+}$ respectively, while the $^3P_1$ level
is uncoupled from the (pseudoscalar) mesonic sector. The $^1P_1$ state does
not couple to $H\bar{H}$ mesons either. Selective coupling results in
different mass shifts of the $Q\bar{Q}$ states, and, eventually, in their
mass splitting. Second, the $H\bar{H}$ continuous spectrum lies above the
quarkonium bound states. Spin-independent potentials reproduce the gross
features of the spectrum accurately enough to assume that the spin-dependent
forces may be treated as a perturbation. The second-order perturbation in energy
\cite{lan} equals
\begin{equation}
E_n^{(2)}=\sum_{m\neq n}\frac{\left| V_{nm}\right| ^2}{E_n^{(0)}-E_m^{(0)}}
\label{ms}
\end{equation}
where $V_{nm}$ is a perturbing interaction and $E_n^{(0)}$ is the
(unperturbed) energy of the $n$-th state. Eq.(\ref{ms}) shows that the mass of the lowest
state $E_0$ (i.e. the quarkonium) decreases because of channel coupling 
(since $E_0^{(0)}<E_m^{(0)}$ ).

Combining these observations we can conclude that the masses of both the $%
^3P_0$ and $^3P_2$ states will diminish because of their coupling to mesonic 
$H\bar{H}$ states, whereas the mass of the $^3P_1$ state will remain
unaffected. Inspection of (\ref{ratio}) shows immediately that the
considered mechanism results in the decrease of the $R$ ratio. 

Next observation is that the magnitude
of the effect depends on the width of the energy gap separating quarkonium
states from the continuous $H\bar{H}$ spectrum: the smaller is the gap, the
more pronounced will be the mass shift. We infer then that the $R(\chi _b)$
ratio ($1^3P_J$ bottomonium states) will be affected less than the $R(\chi
_b^{\prime })$ ratio ($2 ^3P_J$ bottomonium states). Comparing these 
{\it qualitative} predictions to the experimental data (\ref{hie}) we conclude
that the considered mechanism (coupling between quarkonium and mesonic
states) gives a welcome trend to the modified theoretical results.

We proceed now to quantitative estimates of the effect. To this end 
a dynamical model describing the coupling mechanism under
consideration is to developed. We treat quarkonium ($Q\bar{Q}$) and mesonic ($H\bar{H}$)
states as the components of a two-channel system labeling them channels 1
and 2 respectively. Then the Schr\"{o}dinger equations describing the
dynamics of the system for each $J^\pi $ state can be written as follows 
\begin{equation}
\begin{array}{cccc}
T_1\phi _1\;+ & V_{11}\phi _1\;+ & V_{12}\phi _2= & E_1\phi _1 \\ 
T_2\phi _2\;+ &  & V_{21}\phi _1= & E_2\phi _2
\end{array}
\label{sch}
\end{equation}
Here $T$ is the kinetic energy in the corresponding channels and $V$ is the
interaction potential. The $V_{11}$ component is a (linear +\ Coulombic)
potential operating between heavy quarks. Interaction between $H$-mesons is
neglected: near-threshold phenomena under consideration involve small
relative momenta and, hence, large inter-particle distances. At the same
time, the long-range interaction - $\pi -$meson exchange - between two
pseudoscalar particles is absent. Anyway, this approximation does not alter
the qualitative conclusions made above. Finally, $E_1$ and $E_2$ are the
channel energies, differing by the mass gap in the corresponding state: $%
E_2-E_1=2m_H-m_{Q\bar{Q}}$. The non-diagonal component of the potential $%
V_{12}\;(=V_{21})$ describes the dynamics of the quarkonium dissociation $(Q%
\bar{Q})\rightarrow H\bar{H}$, or, stated differently, the dynamics of the $q%
\bar{q}$ production in the field of heavy quarks. Forbidding complexity of
the problem necessitates introducing some simplifying approximations.

Given the large mass difference between heavy and light quarks, we make use
of the adiabatic approximation. The latter molds in the following
assumptions. First, the position of the heavy quarks is not affected by
the process of the $q\bar{q}$ production. Second, the combined spin of heavy
quarks in a pair of heavy mesons is equal to zero. It implies that
in all of the transitions considered the spin of the $Q\bar{Q}$
pair is subject to one and the same change from $S=1$ to $S=0$.  
Finally, c.o.g. of a heavy meson coincides with good accuracy
with the position of the heavy meson residing in it. All the assumptions
made above can be expressed concisely in the following form of the
non-diagonal component: 
\begin{equation}
V_{12}=\beta \;\delta \left( \vec{r}-\vec{r}\sd^{\prime }\right)   
\label{nda}
\end{equation}
where $\beta $ is a coupling strength constant, $\vec{r}$ and 
$\vec{r}\sd^{\prime }$ are the separation between heavy quarks prior and after the $q%
\bar{q}$ production (i.e. within the quarkonium state and in the $H\bar{H}$
pair respectively). The light-quark spin operator and coordinate variables are assumed 
to be present and are dropped for notational brevity.

In the framework of the model the strength constant $\beta $ can be determined 
without going into
details of the complicated mechanism of the light-quark pair production. The
same coupling between channels governs decays of the quarkonium states
lying {\it above }the $H\bar{H}$ threshold, i.e. $\psi (3770)$ and $%
\Upsilon (10580)$ in the case of charmonium and bottomonium respectively.
They are known to decay predominantly into the $D\bar{D}$ and $B\bar{B}$
pairs with the widths equal to 83.9$\pm 2.4$ and 21$\pm 4$ MeV respectively.
The quantum numbers of the bottomonium state are known ($J^\pi =1^{-}$) 
translating into the $4^3S_1$ state of the $b\bar{b}$ pair. The $\psi
(3770)$ quantum numbers are not determined; following the potential model
prescriptions we adopt that it is the $1^3D_1$ state of the $c\bar{c}$
pair \cite{eich,isg}. Using the channel-coupling potential (\ref{nda}) we
calculate the value of the $\beta $
strength coupling constant. It proves to be equal to 153 MeV and 114 MeV for
the charmonium and bottomonium systems respectively. The agreement between
two values of $\beta $ is reasonably good signaling the validity of the
approximations made. Indeed, the paradigm of the non-perturbative QCD
assumes the light-quark pair production as a result of the string break-up.
Flavor-blindness of the confinement forces (materialized as a string)
prompts then that in the infinite heavy-quark mass limit the values of $%
\beta $ for two types of the quarkonium should the same. Given possible
$1/m_Q$ corrections, the $\beta $ values may be
considered as coinciding in line with the theoretical expectation.

Having determined the potential $V_{12}$ connecting the quarkonium and the
mesonic channels we proceed to calculating the matrix element of the
transition between them. The matrix element controls the value of the mass
shift (see (\ref{ms})). Within the approximations made it reduces to the
overlap integral of the quarkonium wave function $\psi _{}^{Q\bar{Q}}$ and
of the wave function $\psi _\nu ^{H\bar{H}}$ of two non-interacting heavy
mesons: 
\begin{equation}
V_\nu =\beta \,\int \psi _{}^{Q\bar{Q}}(\vec{r})\cdot \psi _\nu ^{H\bar{H}}(%
\vec{r})\,d\vec{r}  \label{mel}
\end{equation}
The subscript $\nu $ labels the states of the $H\bar{H}$ continuous
spectrum, to this end we use the c.m.s. momentum of heavy mesons ($\nu
\equiv \vec{k}$). Then the expression for the quarkonium-state mass shift (%
\ref{ms}) can be cast in the form 
\begin{equation}
\Delta m_{Q\bar{Q}}=\int \frac{\left| V_{\vec{k}}\right| ^2}{m_{Q\bar{Q}%
}-m_{H\bar{H}}}\;d\vec{k}  \label{dm}
\end{equation}
where $m_{H\vec{H}}=2m_H+k^2/m_H$. Formal convergence of the integral is
ensured by (slow) decrease of the $V_{\vec{k}}$ matrix element due to
diminishing overlap between the quarkonium bound-state wave function and the
oscillating wave function of two heavy mesons in the continuous spectrum.
However, the integration over $\vec{k}$ which may be translated in the
integration over the mass of the $H\bar{H}$ pair is to be cut-off at much
smaller $m_{\bar{H}H}$ values.
Indeed, the considered mechanism produces mass splitting of the $^3P_J$
states because of the mass difference between the pseudoscalar $H$-meson and
its vector counterpart. We incur then that the scale of integration range
is set by the $m_{H^{*}}-m_H$ mass difference.

Results of numerical calculations confirm the qualitative
conclusions drawn above. To investigate sensitivity of the effect we cut off
the integration in (\ref{dm}) at some $\mu $ ($\mu \geq 2m_H$). Dependence
of the $R$-ratio for the $2P$ and $1P$ states of the bottomonium and for the 
$1P$ states of the charmonium are displayed in fig.1 as a function of the
reduced mass band $r=(\mu/2 - m_H)/(m_{H^{*}}-m_H)$. As the initial values for 
$R$'s we used those calculated in a potential model \cite{grr} (using other
predictions of potential models does not change qualitatively the obtained
results). Note the fast onset of the $R(2P)\sd\lsim\sd R(1P)$ regime: already for 
$r\sd\gsim\sd 0.2$ the hierarchy of the $R$ values corresponds to the experimental
observations (\ref{hie}) and it maintains with further increase of $r$. The
considered mechanism does not yield quantitative agreement with the
experimental data: the overall set of data is shifted downwards with respect
to experimental values (shown by error bars). A possible explanation may be that 
parameters of the
spin-dependent potential determining the "initial" values of $R$ were varied already 
in an attempt to reproduce
experimentally observed regularities. It looks plausible that the consistent
description of spin-dependent mass splitting in the quarkonium systems
requires the account of both the spin-dependent part of the $Q-\bar{Q}$
potential and of the bound-state coupling to mesonic channels. The
investigation of the problem following these lines is in progress.

Summarizing, the process of (virtual) light-quark pair production
considerably affects masses of the quarkonium states. Account of the
emerging mass splitting of the $^3P_J$ triplet allows to reproduce
qualitative features of the experimentally observed quarkonium level spacing.\\[0.3cm]

The author is indebted to F.Lev and S.Romanov for helpful
comments.

\begin{figure}[htb]

\end{figure}
% GNUPLOT: LaTeX picture
\setlength{\unitlength}{0.240900pt}
\ifx\plotpoint\undefined\newsavebox{\plotpoint}\fi
\sbox{\plotpoint}{\rule[-0.200pt]{0.400pt}{0.400pt}}%
\begin{picture}(1500,900)(0,0)
\font\gnuplot=cmr10 at 10pt
\gnuplot
\sbox{\plotpoint}{\rule[-0.200pt]{0.400pt}{0.400pt}}%
\put(220.0,113.0){\rule[-0.200pt]{0.400pt}{184.048pt}}
\put(220.0,113.0){\rule[-0.200pt]{4.818pt}{0.400pt}}
\put(198,113){\makebox(0,0)[r]{0.2}}
\put(1416.0,113.0){\rule[-0.200pt]{4.818pt}{0.400pt}}
\put(220.0,240.0){\rule[-0.200pt]{4.818pt}{0.400pt}}
\put(198,240){\makebox(0,0)[r]{0.3}}
\put(1416.0,240.0){\rule[-0.200pt]{4.818pt}{0.400pt}}
\put(220.0,368.0){\rule[-0.200pt]{4.818pt}{0.400pt}}
\put(198,368){\makebox(0,0)[r]{0.4}}
\put(1416.0,368.0){\rule[-0.200pt]{4.818pt}{0.400pt}}
\put(220.0,495.0){\rule[-0.200pt]{4.818pt}{0.400pt}}
\put(198,495){\makebox(0,0)[r]{0.5}}
\put(1416.0,495.0){\rule[-0.200pt]{4.818pt}{0.400pt}}
\put(220.0,622.0){\rule[-0.200pt]{4.818pt}{0.400pt}}
\put(198,622){\makebox(0,0)[r]{0.6}}
\put(1416.0,622.0){\rule[-0.200pt]{4.818pt}{0.400pt}}
\put(220.0,750.0){\rule[-0.200pt]{4.818pt}{0.400pt}}
\put(198,750){\makebox(0,0)[r]{0.7}}
\put(1416.0,750.0){\rule[-0.200pt]{4.818pt}{0.400pt}}
\put(220.0,877.0){\rule[-0.200pt]{4.818pt}{0.400pt}}
\put(198,877){\makebox(0,0)[r]{0.8}}
\put(1416.0,877.0){\rule[-0.200pt]{4.818pt}{0.400pt}}
\put(220.0,113.0){\rule[-0.200pt]{0.400pt}{4.818pt}}
\put(220,68){\makebox(0,0){0}}
\put(220.0,857.0){\rule[-0.200pt]{0.400pt}{4.818pt}}
\put(463.0,113.0){\rule[-0.200pt]{0.400pt}{4.818pt}}
\put(463,68){\makebox(0,0){0.2}}
\put(463.0,857.0){\rule[-0.200pt]{0.400pt}{4.818pt}}
\put(706.0,113.0){\rule[-0.200pt]{0.400pt}{4.818pt}}
\put(706,68){\makebox(0,0){0.4}}
\put(706.0,857.0){\rule[-0.200pt]{0.400pt}{4.818pt}}
\put(950.0,113.0){\rule[-0.200pt]{0.400pt}{4.818pt}}
\put(950,68){\makebox(0,0){0.6}}
\put(950.0,857.0){\rule[-0.200pt]{0.400pt}{4.818pt}}
\put(1193.0,113.0){\rule[-0.200pt]{0.400pt}{4.818pt}}
\put(1193,68){\makebox(0,0){0.8}}
\put(1193.0,857.0){\rule[-0.200pt]{0.400pt}{4.818pt}}
\put(1436.0,113.0){\rule[-0.200pt]{0.400pt}{4.818pt}}
\put(1436,68){\makebox(0,0){1}}
\put(1436.0,857.0){\rule[-0.200pt]{0.400pt}{4.818pt}}
\put(220.0,113.0){\rule[-0.200pt]{292.934pt}{0.400pt}}
\put(1436.0,113.0){\rule[-0.200pt]{0.400pt}{184.048pt}}
\put(220.0,877.0){\rule[-0.200pt]{292.934pt}{0.400pt}}
\put(45,495){\makebox(0,0){$R_{\chi}$}}
\put(828,23){\makebox(0,0){$r$}}
\put(1071,520){\makebox(0,0)[l]{$R_{\chi_b}$}}
\put(1071,393){\makebox(0,0)[l]{$R_{\chi_b '}$}}
\put(1071,266){\makebox(0,0)[l]{$R_{\chi_c}$}}
\put(1290,706){\makebox(0,0)[l]{$R^{\rm exp}_{\chi_b}$}}
\put(1290,594){\makebox(0,0)[l]{$R^{\rm exp}_{\chi_b '}$}}
\put(1290,495){\makebox(0,0)[l]{$R^{\rm exp}_{\chi_c}$}}
\put(220.0,113.0){\rule[-0.200pt]{0.400pt}{184.048pt}}
\put(1414,704){\raisebox{-.8pt}{\makebox(0,0){$\bullet$}}}
\put(1414,592){\raisebox{-.8pt}{\makebox(0,0){$\bullet$}}}
\put(1414,467){\raisebox{-.8pt}{\makebox(0,0){$\bullet$}}}
\put(1412.0,655.0){\rule[-0.200pt]{0.400pt}{23.608pt}}
\put(1402.0,655.0){\rule[-0.200pt]{4.818pt}{0.400pt}}
\put(1402.0,753.0){\rule[-0.200pt]{4.818pt}{0.400pt}}
\put(1412.0,574.0){\rule[-0.200pt]{0.400pt}{8.672pt}}
\put(1402.0,574.0){\rule[-0.200pt]{4.818pt}{0.400pt}}
\put(1402.0,610.0){\rule[-0.200pt]{4.818pt}{0.400pt}}
\put(1412.0,454.0){\rule[-0.200pt]{0.400pt}{6.263pt}}
\put(1402.0,454.0){\rule[-0.200pt]{4.818pt}{0.400pt}}
\put(1402.0,480.0){\rule[-0.200pt]{4.818pt}{0.400pt}}
\put(220,495){\usebox{\plotpoint}}
\multiput(220.00,493.92)(1.332,-0.498){89}{\rule{1.161pt}{0.120pt}}
\multiput(220.00,494.17)(119.591,-46.000){2}{\rule{0.580pt}{0.400pt}}
\multiput(342.00,447.92)(1.145,-0.498){103}{\rule{1.013pt}{0.120pt}}
\multiput(342.00,448.17)(118.897,-53.000){2}{\rule{0.507pt}{0.400pt}}
\multiput(463.00,394.92)(1.332,-0.498){89}{\rule{1.161pt}{0.120pt}}
\multiput(463.00,395.17)(119.591,-46.000){2}{\rule{0.580pt}{0.400pt}}
\multiput(585.00,348.92)(1.601,-0.498){73}{\rule{1.374pt}{0.120pt}}
\multiput(585.00,349.17)(118.149,-38.000){2}{\rule{0.687pt}{0.400pt}}
\multiput(706.00,310.92)(1.921,-0.497){61}{\rule{1.625pt}{0.120pt}}
\multiput(706.00,311.17)(118.627,-32.000){2}{\rule{0.813pt}{0.400pt}}
\multiput(828.00,278.92)(2.199,-0.497){53}{\rule{1.843pt}{0.120pt}}
\multiput(828.00,279.17)(118.175,-28.000){2}{\rule{0.921pt}{0.400pt}}
\multiput(950.00,250.92)(2.550,-0.496){45}{\rule{2.117pt}{0.120pt}}
\multiput(950.00,251.17)(116.607,-24.000){2}{\rule{1.058pt}{0.400pt}}
\multiput(1071.00,226.92)(2.946,-0.496){39}{\rule{2.424pt}{0.119pt}}
\multiput(1071.00,227.17)(116.969,-21.000){2}{\rule{1.212pt}{0.400pt}}
\multiput(1193.00,205.92)(2.922,-0.496){39}{\rule{2.405pt}{0.119pt}}
\multiput(1193.00,206.17)(116.009,-21.000){2}{\rule{1.202pt}{0.400pt}}
\multiput(1314.00,184.92)(3.448,-0.495){33}{\rule{2.811pt}{0.119pt}}
\multiput(1314.00,185.17)(116.165,-18.000){2}{\rule{1.406pt}{0.400pt}}
\put(220,659){\usebox{\plotpoint}}
\multiput(220.00,657.92)(3.656,-0.495){31}{\rule{2.971pt}{0.119pt}}
\multiput(220.00,658.17)(115.834,-17.000){2}{\rule{1.485pt}{0.400pt}}
\multiput(342.00,640.92)(2.550,-0.496){45}{\rule{2.117pt}{0.120pt}}
\multiput(342.00,641.17)(116.607,-24.000){2}{\rule{1.058pt}{0.400pt}}
\multiput(463.00,616.92)(2.371,-0.497){49}{\rule{1.977pt}{0.120pt}}
\multiput(463.00,617.17)(117.897,-26.000){2}{\rule{0.988pt}{0.400pt}}
\multiput(585.00,590.92)(2.447,-0.497){47}{\rule{2.036pt}{0.120pt}}
\multiput(585.00,591.17)(116.774,-25.000){2}{\rule{1.018pt}{0.400pt}}
\multiput(706.00,565.92)(2.467,-0.497){47}{\rule{2.052pt}{0.120pt}}
\multiput(706.00,566.17)(117.741,-25.000){2}{\rule{1.026pt}{0.400pt}}
\multiput(828.00,540.92)(2.571,-0.496){45}{\rule{2.133pt}{0.120pt}}
\multiput(828.00,541.17)(117.572,-24.000){2}{\rule{1.067pt}{0.400pt}}
\multiput(950.00,516.92)(2.786,-0.496){41}{\rule{2.300pt}{0.120pt}}
\multiput(950.00,517.17)(116.226,-22.000){2}{\rule{1.150pt}{0.400pt}}
\multiput(1071.00,494.92)(2.946,-0.496){39}{\rule{2.424pt}{0.119pt}}
\multiput(1071.00,495.17)(116.969,-21.000){2}{\rule{1.212pt}{0.400pt}}
\multiput(1193.00,473.92)(3.236,-0.495){35}{\rule{2.647pt}{0.119pt}}
\multiput(1193.00,474.17)(115.505,-19.000){2}{\rule{1.324pt}{0.400pt}}
\multiput(1314.00,454.92)(3.448,-0.495){33}{\rule{2.811pt}{0.119pt}}
\multiput(1314.00,455.17)(116.165,-18.000){2}{\rule{1.406pt}{0.400pt}}
\put(220,722){\usebox{\plotpoint}}
\multiput(220.00,720.92)(1.054,-0.499){113}{\rule{0.941pt}{0.120pt}}
\multiput(220.00,721.17)(120.046,-58.000){2}{\rule{0.471pt}{0.400pt}}
\multiput(342.00,662.92)(0.947,-0.499){125}{\rule{0.856pt}{0.120pt}}
\multiput(342.00,663.17)(119.223,-64.000){2}{\rule{0.428pt}{0.400pt}}
\multiput(463.00,598.92)(1.054,-0.499){113}{\rule{0.941pt}{0.120pt}}
\multiput(463.00,599.17)(120.046,-58.000){2}{\rule{0.471pt}{0.400pt}}
\multiput(585.00,540.92)(1.190,-0.498){99}{\rule{1.049pt}{0.120pt}}
\multiput(585.00,541.17)(118.823,-51.000){2}{\rule{0.525pt}{0.400pt}}
\multiput(706.00,489.92)(1.425,-0.498){83}{\rule{1.235pt}{0.120pt}}
\multiput(706.00,490.17)(119.437,-43.000){2}{\rule{0.617pt}{0.400pt}}
\multiput(828.00,446.92)(1.573,-0.498){75}{\rule{1.351pt}{0.120pt}}
\multiput(828.00,447.17)(119.195,-39.000){2}{\rule{0.676pt}{0.400pt}}
\multiput(950.00,407.92)(1.691,-0.498){69}{\rule{1.444pt}{0.120pt}}
\multiput(950.00,408.17)(118.002,-36.000){2}{\rule{0.722pt}{0.400pt}}
\multiput(1071.00,371.92)(1.862,-0.497){63}{\rule{1.579pt}{0.120pt}}
\multiput(1071.00,372.17)(118.723,-33.000){2}{\rule{0.789pt}{0.400pt}}
\multiput(1193.00,338.92)(1.905,-0.497){61}{\rule{1.613pt}{0.120pt}}
\multiput(1193.00,339.17)(117.653,-32.000){2}{\rule{0.806pt}{0.400pt}}
\multiput(1314.00,306.92)(1.921,-0.497){61}{\rule{1.625pt}{0.120pt}}
\multiput(1314.00,307.17)(118.627,-32.000){2}{\rule{0.813pt}{0.400pt}}
\end{picture}
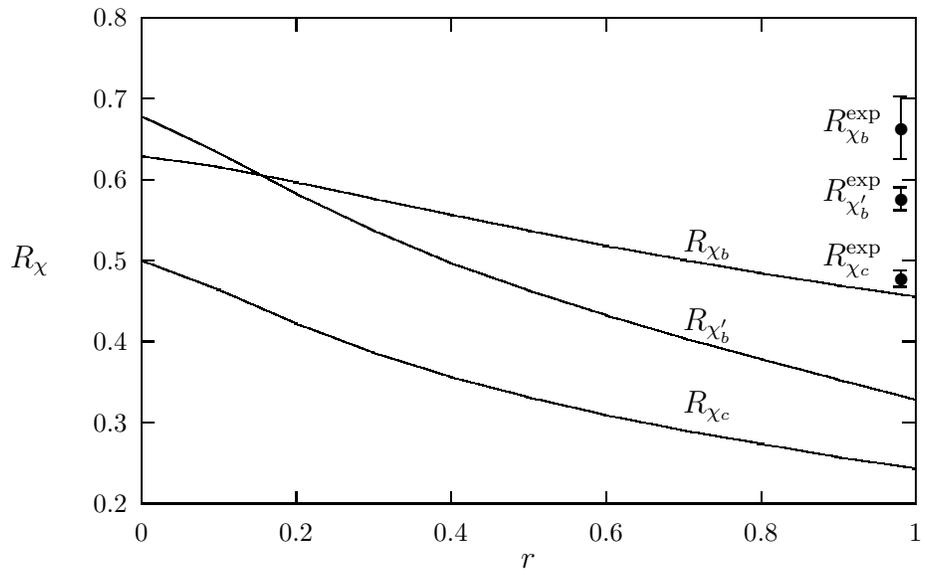
\end{document}